\documentstyle[12pt]{article}
\input{psfig}
\begin{document}

\title{ $f_1$ VECTOR MESON DOMINANCE MODEL FOR THE  STRANGE AXIAL
FORM FACTOR OF THE NUCLEON
\thanks{Contribution to Int.\ Conf.\ on Physics with GeV-Particle Beams,
22.-25.\ 8.\ 1994, J\"ulich, Germany }
}
\author {
M. KIRCHBACH \\
{\it Institut f\"ur Kernphysik, TH Darmstadt,
 D-64289 Darmstadt, Germany}
\and
H. ARENH\"OVEL\\
{\it Institut f\"ur Kernphysik, J. Gutenberg-Universit\"at Mainz}\\
{\it D-55099 Mainz, Germany}
}
\maketitle
\begin{abstract}
The axial strange form factor $F_A^s$ of the nucleon
is assumed to be dominated at low momentum transfer
by the isoscalar axial vector
mesons $f_1(1285)$ and $f_1(1420)$. The importance of
the $a_0\pi N$-triangular
vertex correction is demonstrated.
\end{abstract}

\section{Introduction}

The EMC measurements of the longitudinally
polarized deep inelastic structure function
$g_1^p(x)$ of the proton\cite{1} raised the question of the spin distribution
among quarks and gluons. In the framework of the
``naive'' parton model the proton spin is expressed in terms
of the spin polarization
distribution of quarks and antiquarks
$\Delta q(x) =q^\uparrow (x) -q^\downarrow (x)
+\bar q^\uparrow (x) -\bar q^\downarrow (x)$ as
\begin{equation}
\Delta \Sigma
 =\Delta u +\Delta d +\Delta s\, ,\quad \mbox{ with}\quad
\Delta q =\int_0^1 {\rm d}x \Delta q(x)\, .
\end{equation}
In fact, the EMC has measured the integral
\begin{equation}
\Gamma^p_1 (Q_0^2=10.7\, GeV^2)=
\int_{0.01}^{0.7} dx g_1^p(x, Q_0^2) ={1\over 18}
(4\Delta u +\Delta d + \Delta s)\,
(1- {{\alpha_s(Q_0^2)}\over \pi})\, .
\end{equation}
Note that $\Delta q(Q^2)=$~const.\ in the perturbative regime of
QCD because of chiral symmetry arguments.
In combining (2) with the experimental value of the isovector axial
coupling constant $g_A$ known from neutron $\beta$ decay,
\begin{equation}
g_A^p=-g_A^n =\Delta u -\Delta d\, ,
\end{equation}
on the one side, and with the $F/D$ ratio extracted
from hyperon $\beta$ decay
\begin{equation}
{1\over \sqrt{3}}(3F-D) = {1\over \sqrt{3}}
(\Delta u +\Delta d - 2\Delta s)\, ,
\end{equation}
on the other side, the result is obtained, that the fraction
of the proton spin carried by the quarks is surprisingly small
(see \cite{1a} for details). The integral in (2) can alternatively
be expressed as a combination of the axial isovector ($g_A $),
axial hypercharge ($G_A^{(8)}$), and axial flavor singlet ($G_A^{(0)}$)
form factors as follows:
\begin{eqnarray}
\int_0^1 {\rm d}x\, g_1^p(x) & = &
{1\over 12\sqrt{3}} (\sqrt{3}g_A +2 G_A^{(8)} +
4\sqrt{2}G_A^{(0)})\,(1-{{\alpha_s(Q^2)}\over \pi})\nonumber\\
& = & {1\over 6}({g_A\over 2} +{5\over {\sqrt{3}} }G_A^{(8)} +2G_A^s)
(1-{{\alpha_s(Q^2)}\over \pi}) \, ,
\end{eqnarray}
where (5) is based on the (presumably rather strong)
assumption of SU(3) symmetry leading to the relations
\begin{equation}
G_A^{s}  =\sqrt{{2\over 3}}
G_A^{(0)}-2\sqrt{{1\over 3}}G_A^{(8)}\, ,\qquad
{2\over \sqrt{3}}G_A^{(8)}  = {1\over 3}(3F -D)\, .
\end{equation}
The nucleon matrix elements of
the axial hypercharge ($A_\mu^{(8)}$) and flavor singlet ($A_\mu^{(0)}$)
quark currents are normalized according to:
\begin{equation}
\langle N\mid A_\mu^{(0, 8)} \mid N\rangle   =
\langle N\mid \bar{\Psi}\gamma_\mu\gamma_5 {{\lambda^{(0,8)}}\over 2}
\Psi\mid N\rangle  =
G_A^{(0,8)}
\bar{{\cal U}}\gamma_\mu\gamma_5 {\cal U}\, ,
\end{equation}
where  $\lambda^{(0,8)}$ denote the corresponding Gell-Mann matrices,
${\cal U}$ the nucleon Dirac spinor\cite{2},
and $\Psi $ stands for the flavor quark SU(3) spinor.
The strange axial nucleon form factor $G_A^{s}$ is introduced as
\begin{equation}
\langle N\mid \bar{s}\, \gamma_\mu\gamma_5\, s \mid N\rangle   =
G_A^{s} \bar{{\cal U}} \gamma_\mu\gamma_5{\cal U}\, .
\end{equation}

In the case of a vanishing strange axial form factor, eq.\ (5) reduces
to the so called Ellis-Jaffe sum rule\cite{3}.
Now the surprise is the significant deviation
of the measured value for $\Gamma_1^p$ from
the Ellis-Jaffe sum rule, attributed to
a small flavor
singlet axial form factor or, alternatively, to
a significant nonvanishing value of $G_A^s$.
Different
scenarios have been discussed in the literature in this respect.
Here, we would like to point out the one presented in\cite{4} where
due to a non-trival structure of the
hadronic vacuum
the flavor singlet axial charge is ``eaten'' by instanton effects.
The axial form factors introduced above enter the neutral
axial vector current according to (for details see, e.g., ref.\ \cite{5})
\begin{eqnarray}
\langle N\mid A_\mu\mid N \rangle  & = &
\langle N\mid - \bar{\psi}\gamma_\mu\gamma_5 {\tau_3\over 2} \psi
              +{1\over 4}\bar{s}\, \gamma_\mu\gamma_5\, s
              +{1\over 4} \bar{c}\, \gamma_\mu\gamma_5\, c\mid N\rangle
\nonumber\\
& = & -{g_A\over 2} \bar {{\cal U}}\gamma_\mu\gamma_5\tau_3\, {\cal U}
+{G_A^s\over 4} \bar{{\cal U}}\gamma_\mu\gamma_5{\cal U}\, .
\end{eqnarray}
Here, the notation $\bar{\psi} =(u,d)$
has been used and the $\bar{c}c$
content of the nucleon has been neglected in the last line of (9).
Thus in the light of the proton spin problem, the theoretical
investigation of the strange axial nucleon form factor becomes
important.

In this study we present a $f_1$-vector meson dominance (VMD)
model for this form factor and demonstrate the importance of
a lowest 1-loop vertex correction.

\section{The Structure of the $f_1NN$ Vertex}

In the spirit of the VMD ansatz\cite{6}  and
in analogy to the $A_1$ dominance of the isotriplet
axial current exploited in\cite{7}
we suggest the following representation for
the strange axial form factor of the nucleon at low energies:
\begin{equation}
F^s_A(t) =\frac{G^s_A}{\bar g}
\left\{- {{g_{D}m_D^2}\over {m_D^2 -t}}
                            {F_D(t)}\sin \epsilon
                          + {{g_{E}m_E^2}\over {m_E^2 -t}}
                            {F_E(t)} \cos \epsilon\right\}\, .
                           \end{equation}
where $\bar g=g_{E}\cos \epsilon - g_{D}\sin \epsilon$.
Here, ``$D$'' and ``$E$'' label respectively the constants related to the
$f_1(1285)$ and $f_1(1420)$ mesons. In (10) the deviation
of the mixing angle between the isoscalar of the octet
and the flavor singlet state in the axial vector meson nonet
from the ideal mixing
angle of $\theta_0 = 35.3^\circ$ has been made explicit according to:
\begin{eqnarray}
f_1(1285) & = &
{ {\bar{u}u+\bar{d}d}\over {\sqrt{2}}}   \cos \epsilon \,
-\bar{s}s \, \sin \epsilon\, , \\
f_1(1420) & = &
{ {\bar{u}u+\bar{d}d} \over {\sqrt{2}}} \, \sin  \epsilon\,
+ \bar{s}s \, \cos \epsilon\,.
\end{eqnarray}
The angle $\epsilon$ has been determined  experimentally in\cite{8}
to be $\epsilon  = 15^\circ \pm _{5^\circ}^{10^\circ} $.
The latter equation shows that a significant violation of the OZI rule
is observed in the axial vector nonet\cite{9}.
By chiral symmetry arguments it is possible to parametrize the coupling
constants of the $f_1$ mesons as
$ g_{D}=g_\omega \cos \epsilon -g_\phi \sin \epsilon$ and
$g_{E}=g_\omega \sin \epsilon +g_\phi \cos \epsilon$. The values for
$g_\omega$ and $g_\phi$ are taken from\cite{Ge76}.

Dispersion relations with one subtraction have now been
assumed for the
$f_1(1285)NN$ and $f_1(1420)NN$ strange form factors, respectively,
\begin{eqnarray}
F_{D/E}(t) & = &
1 + {t\over \pi }
\int_{{\rm threshold}}^\infty {\rm d}t'
{{\Im (F_{D/E}(t'))}\over {t'(t'-t-i\epsilon )}}\, .
\end{eqnarray}

Now, the imaginary part of $F_{D/E}(t)$ can be calculated from the Feynman
graphs of Fig.~1 by means of the Cutkosky rules\cite{10}.
The most important $f_1NN$ vertex correction is due to the
($f_1(1285)\to a_0(980) +\pi$) decay channel with a $ (37\pm 7)\% $
branching ratio. It is especially
important for the modification of the one-body isoscalar
axial vector current of in-medium nucleons, a fact demonstrated
in\cite{11}.
There it was shown for the case of the Fermi gas model that
the $a_0\pi$ exchange
current  reduces the one-body matrix element of the
isoscalar axial vector current
by about 10\%. The effect is determined by
the ratio $\sin \epsilon /(g_{D}\, G_A^s)$ and acquires importance
mainly because of the above mentioned significant violation of the
OZI rule in the axial vector meson nonet. We use the value $G_1^s=
\frac{1}{2}G_A^s=-0.13\pm 0.04$ in accordance with\cite{1a}.

\section{Results and Discussion}

To describe the $f_1(1285) a_0\pi$ vertex we use the following
effective Lagrangian density
\begin{equation}
{\cal L}_{Da_0\pi} = g_{D a_0\pi}f_1^\lambda \partial_\lambda
                       \vec{\phi}_\pi\cdot\vec{a}_0\, .
\end{equation}
The value of $g_{Da_0\pi}$ extracted from the corresponding
partial width is up to a sign $|g_{Da_0\pi}| =5.3$. The coupling constant
of the $a_0(980)$ meson to the nucleon is identified
to that of the scalar-isotriplet channel of the Bonn potential\cite{12}
as $g_{a_0NN} =3.73 $, and
a pseudoscalar form of the $\pi N$ vertex is assumed.
{}From the graphs in Fig.~1 we obtain for the
absorptive part of the $f_1NN$ form factor  the following expressions
for its low frequency part ($(m_{a_0} +m_\pi)^2\leq t\leq 4m_N^2$):
\begin{equation}
\Im (F_D(t))  =
c\,{{m_N t-(m_{a_0}^2 - m_\pi^2)(\sqrt{t}-m_N)}
\over {t\sqrt{\mid t -4m_N^2\mid }}}
\arctan  A(t)\, ,\label{im}
\end{equation}
where
\begin{eqnarray}
A(t) & =& { \sqrt{\mid t -4m_N^2\mid }\sqrt{t-2(m_{a_0}^2 +m_\pi^2)
         +(m_{a_0}^2 -m_\pi^2)^2/t}\over {t-m_{a_0}^2 -m_{\pi}^2}}\, ,
\nonumber\\
c & =& { {g_{Da_0\pi}g_{a_0NN}g_{\pi NN}}\over {8\pi g_{D}G_A^s}} \, .
 \nonumber
\end{eqnarray}
The high frequency part ($4m_N^2\leq t\leq \infty$) is simply obtained
from (\ref{im}) by replacing $\arctan  A(t)$ by
${1\over 2}\ln {{1-A(t)}\over {1+A(t)}}$. It is small for unitarity arguments.

In Fig.\ 2 the ratio of $F^s_A(t)$ including the vertex contribution to the
one ($F^s_{A,0}(t)$) without is presented. At higher momentum transfers, the
vertex contribution is on the level of several percent and thus non negligible.
We expect that the $a_0\pi N$ triangular
vertex correction to the strange axial form factor
will increase in case the $a_0\pi $ interaction is included.
The relevant T-matrix amplitude
can be evaluated within the framework presented in\cite{13}.
The main conclusion of our study is that
the strong violation of the OZI
rule in the axial vector meson nonet implies

This work has been partially supported by Gesellschaft f\"ur
Schwerionen\-for\-schung (Darmstadt) and Deutsche Forschungsgemeinschaft (SFB
201).

\vspace{.5cm}
\centerline{\psfig{figure=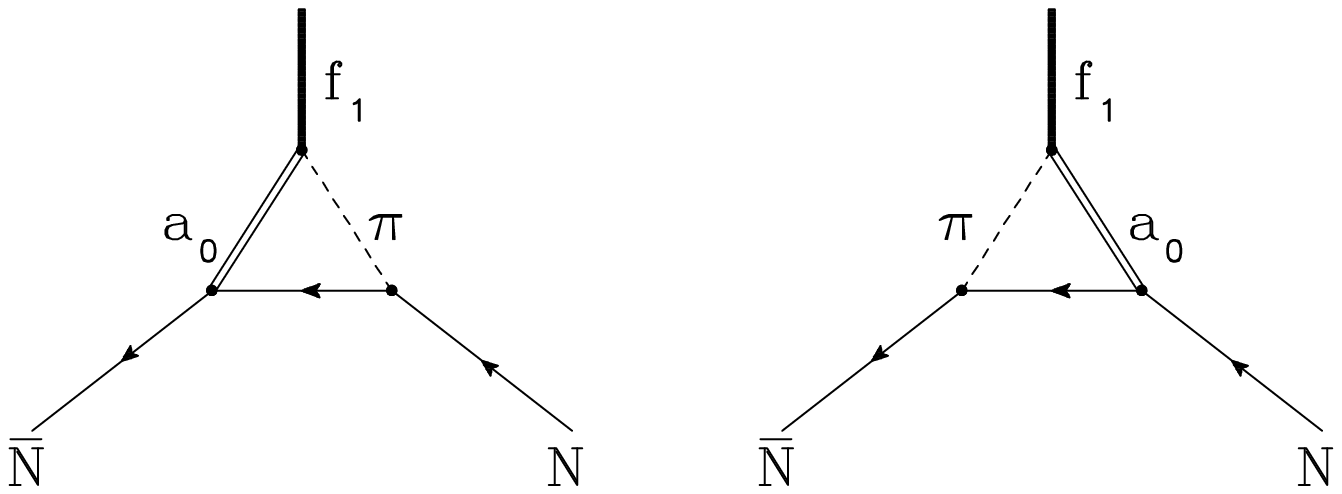,width=10cm,angle=90}}

\centerline{{\small\noindent{\bf Fig.~1}:
Triangle vertex correction to the strange axial form factor of the
nucleon.}}

\vspace{0.4cm}
\centerline{\psfig{figure=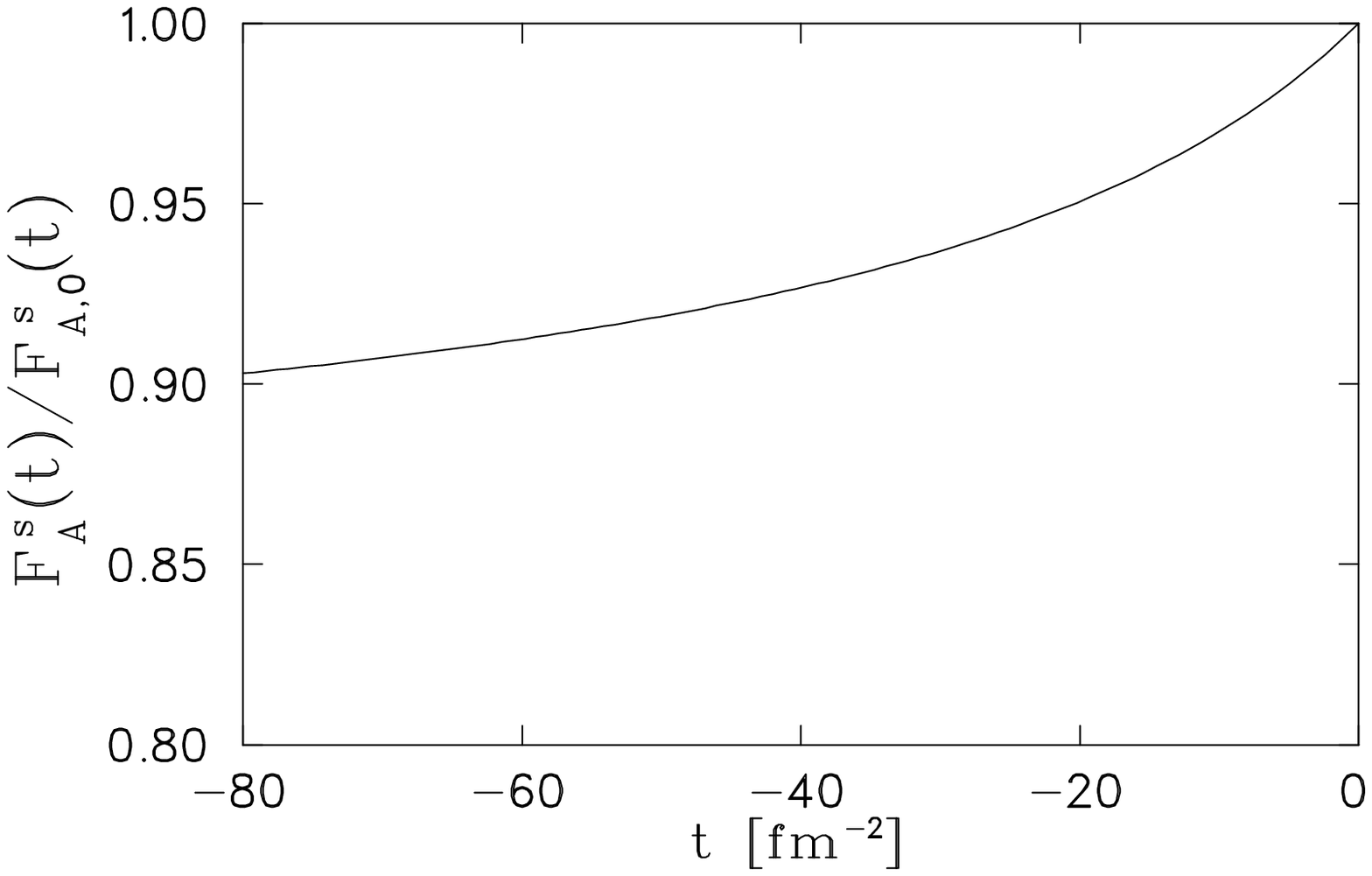,width=10cm,angle=0}}

\centerline{{\small\noindent{\bf Fig.~2}:
Ratio of the strange axial form factor with and without vertex correction.}}
\vspace{0.5cm}

\end{document}